\documentclass[11pt]{article}
\usepackage{amsmath, amssymb, latexsym, amscd, amsfonts ,amstext}
\usepackage[mathscr]{eucal}
\usepackage{graphicx}
\usepackage{subfig}

\date{}

\input{tcilatex}
\begin{document}

\title{ Uniqueness of a phaseless inverse scattering problem for the
generalized 3-D Helmholtz equation \thanks{%
This work was supported by US Army Research Laboratory and US Army Research
Office grant W911NF-15-1-0233 as well as by the Office of Naval Research
grant N00014-15-1-2330. }}
\author{Michael V. Klibanov\thanks{
Department of Mathematics and Statistics, University of North Carolina at
Charlotte, Charlotte, NC 28223, USA, E-mail: \texttt{mklibanv{@}uncc.edu}.}}
\maketitle

\begin{abstract}
An inverse scattering problems for the 3-D generalized Helmholtz equation is
considered. Only the modulus of the complex valued scattered wave field is
assumed to be measured and the phase is not measured. Uniqueness theorem is
proved.
\end{abstract}


\textbf{Keywords}: phaseless inverse scattering, generalized Helmholtz equation, uniqueness

\textbf{2010 Mathematics Subject Classification}: 35R30

\pagestyle{myheadings} \thispagestyle{plain} 
\markboth{M.V.~Klibanov and
V.G. Romanov}{Inverse scattering problems without the phase information}

\graphicspath{{Figures/}}

%
%
%
%
%
%
%
%
%
%
%

%
%

\graphicspath{{FIGURES/}
{Figures/}
{FiguresJ/newfigures/}
{pics/}}

\section{Introduction}

\label{sec:1}

Phaseless Inverse Scattering Problems (PISPs) arise in imaging of
nanostructures \cite{Dar,Die,Pet,Ruhl} and biological cells \cite{PM,Bio}.
In these applications, sizes of objects of interest are typically on the
micron scale or less. Recall that $1micron=1\mu m=10^{-6}m$ where
\textquotedblleft $m$" stands for meter. Sizes of many nanostructures are
usually hundreds of nanometers (nm), $100nm=10^{-7}m=0.1\mu m.$ Sizes of
biological cells are in the range of $\left( 5,100\right) \mu m$ \cite%
{PM,Bio}. Therefore, in imaging of these objects the wavelengths of the
electromagnetic signals should also be in the same range. As to those
signals, usually either X-rays or optical signals are used. It is well known
that the phase of an electromagnetic signal cannot be measured for such
small wavelengths \cite{Dar,Die,Pet,Ruhl}. Therefore, we arrive at the
problem of the reconstruction of the spatially distributed dielectric
constant of a scatterer using only the intensity of the scattered wave
field. The intensity is the square modulus of the complex valued wave field.

A similar problem, although for the Schr\"{o}dinger equation in the
frequency domain, arises in the quantum inverse scattering, where only the
differential scattering cross-section can be measured, i.e. the square
modulus of the solution of that equation, see, e.g. page 8 of \cite{Newton}
and Chapter 10 of \cite{CS}. Unlike the generalized Helmholtz equation
considered in this paper, in the case of the Schr\"{o}dinger equation the
unknown potential is not multiplied by $k^{2},$ which significantly
simplifies the problem. Note that, unlike PISPs, the conventional inverse
scattering theory is based on the assumption that both the intensity and the
phase of the complex valued wave field are measured, see, e.g. \cite%
{Bao3,CS,CK,Hu,Is,Li,Newton,Nov1,Nov2}.\ Here and below $k$ is the wave
number.

Let $\Omega ,G\subset \mathbb{R}^{3}$ be two bounded domains and $\Omega
\subset G$. Let $S=\partial G,S\in C^{1}$ and $S\cap \partial \Omega
=\varnothing .$ Let $\omega >0$ be \ a number. For every $y\in \mathbb{R}%
^{3} $ denote $B_{\omega }\left( y\right) =\left\{ x\in \mathbb{R}%
^{3}:\left\vert x-y\right\vert <\omega \right\} $ the ball of the radius $%
\omega $ with the center at the point $y$. Below $c(x),x\in \mathbb{R}^{3}$
is the spatially distributed dielectric constant. We assume that $c(x)$ is a
real valued function satisfying the following conditions 
\begin{equation}
c\in C^{15}(\mathbb{R}^{3}),\quad c(x)=1+\beta \left( x\right) ,  \label{1.1}
\end{equation}%
\begin{equation}
\beta (x)\geq 0,\text{ }\beta \left( x\right) =0\quad \text{for }\>x\in 
\mathbb{R}^{3}\setminus \Omega .  \label{1.3}
\end{equation}%
The smoothness requirement imposed on the function $c(x)$ is due to Lemma
2.1 (below) as well as due to Theorem 2 in \cite{KR1}. The conformal
Riemannian metric generated by the function $c(x)$ is 
\begin{equation}
d\tau =\sqrt{c\left( x\right) }\left\vert dx\right\vert ,|dx|=\sqrt{%
(dx_{1})^{2}+(dx_{2})^{2}+(dx_{3})^{2}}.  \label{1.31}
\end{equation}%
\ Below we rely on the following Assumption:

\textbf{Assumption}. \emph{We assume that geodesic lines of the metric (\ref%
{1.31}) satisfy the regularity condition, i.e. for each two points }$x,y\in 
\mathbb{R}^{3}$\emph{\ there exists a single geodesic line }$\Gamma \left(
x,y\right) $\emph{\ connecting these points.}

It is well known from the Hadamard-Cartan theorem \cite{Ball} that in any
simply connected complete manifold with a non positive curvature each two
points can be connected by a single geodesic line. The manifold $(\Omega
,\varepsilon _{r})$ is called the manifold of a non positive curvature, if
the section curvatures $K(x,\sigma )\leq 0$ for all $x\in \overline{\Omega }$
and for all two-dimensional planes $\sigma $. A sufficient condition for the
inequality $K(x,\sigma )\leq 0$ was derived in \cite{R4}. This condition is: 
\begin{equation}
\sum_{i,j=1}^{3}\frac{\partial ^{2}\ln c(x)}{\partial x_{i}\partial x_{j}}%
\xi _{i}\xi _{j}\geq 0,\>\forall \xi \in \mathbb{R}^{3},\forall x\in 
\overline{\Omega }.  \label{1.310}
\end{equation}%
Thus, (\ref{1.310}) is a sufficient condition, which guarantees the validity
of Assumption in terms of the function $c(x).$ For $x,y\in \mathbb{R}^{3},$
let $\tau (x,y)$ be the solution to the following problem: 
\begin{equation}
|\nabla _{x}\tau (x,y)|^{2}=c(x),\quad  \label{1.9}
\end{equation}%
\begin{equation}
\tau (x,y)=O\left( \left\vert x-y\right\vert \right) \text{ }\>\mathrm{as}%
\text{ }\>x\rightarrow y.  \label{1.900}
\end{equation}%
Equation (\ref{1.9}) is called \textquotedblleft eikonal equation". Let $%
d\sigma $ be the euclidean arc length of the geodesic line $\Gamma \left(
x,y\right) .$ Then the solution of the problem (\ref{1.9}), (\ref{1.900}) is 
\begin{equation}
\tau \left( x,y\right) =\dint\limits_{\Gamma \left( x,y\right) }c\left( \xi
\right) d\sigma .  \label{1.90}
\end{equation}%
Hence, $\tau (x,y)$ is the travel time between points $x$ and $y$ due to the
Riemannian metric (\ref{1.31}). Due to the Assumption, $\tau (x,y)$ is a
single-valued function of both points $x$ and $y$ in $\mathbb{R}^{3}\times 
\mathbb{R}^{3}$.

We consider the following equation 
\begin{equation}
\Delta u+k^{2}c(x)u=-\delta (x-y),\quad x\in \mathbb{R}^{3},  \label{1.4}
\end{equation}%
where the Laplace operator is taken with respect to $x$, the wave number $%
k>0 $ and $y\in \mathbb{R}^{3}$ is the source position. Naturally, we assume
that the function $u(x,y,k)$ satisfies the radiation condition 
\begin{equation}
\frac{\partial u}{\partial r}-iku=o(r^{-1})\>\text{ }\mathrm{as}\text{ }%
\>r=|x-y|\rightarrow \infty .  \label{1.5}
\end{equation}%
Denote $u_{0}(x,y,k)$ the solution of the problem (\ref{1.4}), (\ref{1.5})
for the case $n(x)\equiv 1.$ Then $u_{0}$ is the incident spherical wave,%
\begin{equation}
u_{0}(x,y,k)=\frac{\exp \left( ik\left\vert x-y\right\vert \right) }{4\pi
\left\vert x-y\right\vert }.  \label{1.50}
\end{equation}%
Let $u_{sc}(x,y,k)$ be the scattered wave, which is due to the presence of
scatterers, in which $c(x)\neq 1$. Then 
\begin{equation}
u_{sc}(x,y,k)=u(x,y,k)-u_{0}(x,y,k)=u(x,y,k)-\frac{\exp \left( ik\left\vert
x-y\right\vert \right) }{4\pi \left\vert x-y\right\vert }.  \label{1.6}
\end{equation}%
Combining Theorem 8.7 of \cite{CK} with Theorem 6.17 of \cite{GT} and taking
into account that $c\in C^{15}(\mathbb{R}^{3}),$ we obtain that the problem (%
\ref{1.4}), (\ref{1.5}) has unique solution

$u\in C^{16+\alpha }\left( \left\vert x-x_{0}\right\vert \geq \varepsilon
\right) ,\forall \varepsilon >0$ for any $\alpha \in \left( 0,1\right) .$
Here $C^{16+\alpha }$ is the H\"{o}lder space.

We model the propagation of the electric field by a single equation (\ref%
{1.4}) since it was demonstrated numerically in \cite{BMM} that this is
possible in the case when only one component of this field is incident upon
the medium. Indeed, it was shown in \cite{BMM} that in this case that
component dominates two other components and also its propagation is well
governed by a single PDE, which is a direct analog of (\ref{1.4}). This
conclusion was verified in \cite{BK,TBKF1,TBKF2} via accurate imaging from
experimental data.

Denote $dist\left( S,\partial \Omega \right) >0$ the Hausdorf distance
between the surface $S$ and the domain $\Omega .$

\textbf{Phaseless Inverse Scattering Problem (PISP)}. Let the number $\omega
\in \left( 0,dist\left( S,\partial \Omega \right) \right) .$ Let $u(x,y,k)$
be the solution of the problem (\ref{1.4}), (\ref{1.5}). Assume that the
following function $f\left( x,y,k\right) $ is known%
\begin{equation}
f\left( x,y,k\right) =\left\vert u\left( x,y,k\right) \right\vert ,\forall
y\in S,\forall x\in B_{\omega }\left( y\right) ,x\neq y,\forall k\in \left(
a,b\right) ,  \label{1.7}
\end{equation}%
where $\left( a,b\right) \subset \left\{ z>0\right\} $ is a certain
interval. Determine the function $c\left( x\right) .$

Theorem 1 is the main results of this paper. Compared with previous results 
\cite{KSIAP,AML,AA}, the main difficulty in the proof of this theorem is due
to the necessity of using the apparatus of the Riemannian geometry.

\textbf{Theorem 1}. \emph{Consider an arbitrary pair of points }$y\in S,x\in
B_{\omega }\left( y\right) ,x\neq y.$\emph{\ And consider the function }$%
g_{x,y}\left( k\right) =f\left( x,y,k\right) $ \emph{as the function of the
variable }$k,$\emph{\ where the function }$f\left( x,y,k\right) $\emph{\ is
defined in (\ref{1.7}). Then the function }$\varphi _{x,y}\left( k\right)
=u\left( x,y,k\right) $\emph{\ of the variable }$k$\emph{\ is reconstructed
uniquely, as soon as the function }$g_{x,y}\left( k\right) $\emph{\ is given
for all }$k\in \left( a,b\right) .$\emph{\ The PISP has at most one
solution. }

The first uniqueness theorem for a PISP for a 1-D Schr\"{o}dinger equation
was proved in \cite{KS}. More recently uniqueness theorems were proved for
3-D PISPs for the Schr\"{o}dinger equation in \cite{KSIAP,AML}. Also, in 
\cite{AA} uniqueness was proved for the 3-D PISP for the generalized
Helmholtz equation in the case when the function $\delta (x-y)$ in the right
hand side of (\ref{1.4}) was replaced by a function $p\left( x\right) $ such
that 
\begin{equation}
p\left( x\right) \neq 0\text{ in }\overline{\Omega }.  \label{1.82}
\end{equation}

Our PISP is overdetermined. Indeed, the unknown coefficient $c\left(
x\right) $ depends on three variables, whereas the data $f$ depend on six
variables. On the other hand, even if the phase is known, currently there
are no uniqueness theorems, which would be proven for a non-overdetermined
statement of a 3-D inverse scattering problem in the case when the $\delta -$%
function is in the right hand side of (\ref{1.4}). On the other hand the
PISP in \cite{AA} is non overdetermined. In fact, that is an inverse problem
with the data generated by a single measurement event. However, the price to
pay in \cite{AA}, is the assumption (\ref{1.82}). The same is true for the
PISP for the Schr\"{o}dinger equation on page 397 of \cite{KSIAP}.

Reconstruction procedures for PISPs both for the generalized Helmholtz
equation and for the Schr\"{o}dinger equation were developed in \cite%
{KR1,KR2,KR,KRB}. In \cite{KNP} an essentially modified procedure of \cite%
{KR1} was implemented numerically. In \cite{Nov3,Nov4} different statements
of PISPs were proposed, which led to different uniqueness theorems and
reconstruction procedures. While we are interested in the reconstruction of
the unknown coefficient of a PDE, we also refer to \cite%
{Am,Bao1,Bao2,Iv1,Iv2} for numerical solutions of PISPs in the case when the
surface of a scatterer was reconstructed.

Everywhere below we assume that conditions of Theorem 1 as well as
Assumption hold. In section 2 we prove some lemmata. In section 3 we prove
Theorem 1.

\section{Lemmata}

\label{sec:2}

For any complex number $z\in \mathbb{C}$ we denote by $\overline{z}$ its
complex conjugate. Let $\gamma >0$ be a number. Denote%
\begin{equation*}
\mathbb{C}_{\gamma }=\left\{ z\in \mathbb{C}:\func{Im}z>-\gamma \right\} ,%
\mathbb{C}_{+}=\left\{ z\in \mathbb{C}:\func{Im}z>0\right\} .
\end{equation*}%
Hence, $\mathbb{C}_{+}$ is the upper half plane of the complex plane $%
\mathbb{C}.$

\subsection{Some properties of the function $u(x,y,k)$}

\label{sec:2.1}

Let $\zeta =(\zeta _{1},\zeta _{2},\zeta _{3})$, $\zeta =\zeta (x,y)$ be
geodesic coordinates of a variable point $x$ with respect to a fixed point $%
y $ in the above Riemannian metric (\ref{1.31}). By the Assumption, there
exists a one-to-one correspondence $x\Leftrightarrow \zeta $ for any fixed $%
y $. Consider the Jacobian $J(x,y)$, 
\begin{equation}
J(x,y)=\mathrm{\det }\frac{\partial \zeta }{\partial x}.  \label{1.91}
\end{equation}%
It was proven in \cite{KR1} that $J(x,y)>0$ for all $x,y$. Let $T>0$ be an
arbitrary number. Consider the domains $D\left( y,T\right) $ and $D^{\ast
}\left( y,T\right) $ defined as 
\begin{equation*}
D(y,T)=\{(x,t):0<\>t\leq T-\tau (x,y)\},
\end{equation*}%
\begin{equation*}
D^{\ast }(y,T)=\{(x,t):\tau (x,y)\leq t\leq T-\tau (x,y)\}.
\end{equation*}

\textbf{Lemma 2.1}.\emph{\ Let }$y\in \mathbb{R}^{3}$\emph{\ be an arbitrary
point. Then there exists a number }$\gamma =\gamma \left( y,G\right) >0$%
\emph{\ such that for every }$x\in G$\emph{\ the function }$u\left(
x,y,k\right) $\emph{\ can be analytically continued with respect to }$k$%
\emph{\ from the half real line }$\mathbb{R}_{+}=\left\{ k:k>0\right\} $%
\emph{\ in the half plane }$\mathbb{C}_{\gamma }.$

\textbf{Proof}. Consider the following associated Cauchy problem%
\begin{equation}
c\left( x\right) v_{tt}=\Delta v+\delta \left( x-y,t\right) ,x\in \mathbb{R}%
^{3},t>0,  \label{2.1}
\end{equation}%
\begin{equation}
v\mid _{t<0}\equiv 0.  \label{2.2}
\end{equation}%
It was proven in \cite{KR1} that the solution of this problem can be
represented as%
\begin{equation}
v(x,y,t)=A(x,y)\delta (t-\tau (x,y))+\hat{v}(x,y,t)H(t-\tau (x,y)),
\label{2.3}
\end{equation}%
where the function $\hat{v}(x,y,t)\in \in C^{2}\left( D^{\ast }(y,T)\right)
, $ 
\begin{equation}
A(x,y)=\frac{\sqrt{J(x,y)}}{4\pi n(x)\tau (x,y)}  \label{2.30}
\end{equation}%
and $H\left( t\right) $ is the Heaviside function,%
\begin{equation*}
H\left( t\right) =\left\{ 
\begin{array}{c}
1,t>0, \\ 
0,t<0.%
\end{array}%
\right.
\end{equation*}

It follows from \ Lemma 6 of Chapter 10 of the book \cite{V} as well as
Remark 3 after that lemma that for any fixed point $y\in \mathbb{R}^{3}$ the
function $v(x,y,t)$ decays exponentially as $t\rightarrow \infty $ together
with its $x-$derivatives up to the second order. This decay is uniform for
all $x\in \overline{G}.$ In other words, there exist constants $M=M\left(
G,c\right) >0,m=m\left( G,c\right) >0$ such that 
\begin{equation}
\left\vert D_{x,t}^{\alpha }v\left( x,y,t\right) \right\vert \leq
Me^{-m\,t},\forall \text{ }\>t\geq t_{0}\>,\forall \>x\in \overline{G},
\label{2.4}
\end{equation}%
where $t_{0}=t_{0}\left( G,c\right) =const.>0.$ Here $\alpha =\left( \alpha
_{1},\alpha _{2},\alpha _{3},\alpha _{4}\right) $ is the multiindex with
non-negative integer coordinates and $\left\vert \alpha \right\vert =\alpha
_{1}+\alpha _{2}+\alpha _{3}+\alpha _{4}\leq 2.$ Hence, one can consider the
Fourier transform $W\left( x,y,k\right) $ of the function $v,$%
\begin{equation}
W(x,y,k)=\dint\limits_{0}^{\infty }v\left( x,y,t\right) \exp \left(
ikt\right) dt.  \label{2.5}
\end{equation}%
Next, theorem 3.3 of \cite{V1} and theorem 6 of Chapter 9 of \cite{V}
guarantee that 
\begin{equation}
W(x,y,k)=u(x,y,k),  \label{2.50}
\end{equation}%
where the function $u(x,y,k)$ is the above solution of the problem (\ref{1.4}%
), (\ref{1.5}).

Hence, it follows from (\ref{2.4}) and (\ref{2.5}) that the function $%
u(x,y,k)$ can be analytically continued from $\mathbb{R}_{+}$ in the half
plane $\mathbb{C}_{m}.$ $\square $

An analog of Lemma 2.2 was proven in \cite{KR1} for the case when $k$ is a
real number.

\textbf{Lemma 2.2}. \emph{Let }$A(x,y)$\emph{\ be the function defined in (%
\ref{2.30}). The asymptotic behavior of the function }$u(x,y,k)$\emph{\ is}%
\begin{equation}
u(x,y,k)=A(x,y)e^{ik\tau \left( x,y\right) }\left( 1+O\left( \frac{1}{k}%
\right) \right) ,\left\vert k\right\vert \rightarrow \infty ,k\in \mathbb{C}%
_{m},x\in \overline{G}.  \label{2.6}
\end{equation}

Proof. By (\ref{2.3}) and (\ref{2.5})%
\begin{equation}
u(x,y,k)=A(x,y)e^{ik\tau \left( x,y\right) }+\dint\limits_{\tau \left(
x,y\right) }^{\infty }\hat{v}(x,y,t)\exp \left( ikt\right) dt.  \label{2.7}
\end{equation}%
The integration by parts in (\ref{2.7}) and (\ref{2.4}) lead to (\ref{2.6}). 
$\square $

\subsection{Four more lemmata}

\label{sec:2.2}

\textbf{Lemma 2}.\textbf{3}. \emph{For any pair }$x\in \overline{G},y\in 
\mathbb{R}^{3},x\neq y$\emph{\ the function }$u(x,y,k)$\emph{\ has at most
finite number of zeros in }$\mathbb{C}_{m}.$

\textbf{Proof}. Follows immediately from (\ref{2.6}). $\square $

\textbf{Lemma 2.4.} \emph{Let the function }$r\left( k\right) $\emph{\ be
analytic in the half plane }$\mathbb{C}_{m}$\emph{\ and has no zeros in }$%
\mathbb{C}_{+}\cup \mathbb{R}.$\emph{\ Assume that } 
\begin{equation*}
r\left( k\right) =\frac{C}{k^{n}}\left[ 1+o\left( 1\right) \right] \exp
\left( ikL\right) ,\left\vert k\right\vert \rightarrow \infty ,k\in \mathbb{C%
}_{+},
\end{equation*}%
\emph{where }$C\in \mathbb{C}$\emph{, }$n\emph{,}L\in \mathbb{R}$\emph{\ are
some numbers and also }$n\geq 0$\emph{. Then the function }$r\left( k\right) 
$\emph{\ can be uniquely determined for }$k\in \mathbb{C}_{+}\cup \mathbb{R}$%
\emph{\ by the values of }$\left\vert r\left( k\right) \right\vert $\emph{\
for }$k\in \mathbb{R}$\emph{. }

Lemma 2.4 follows immediately from Proposition 4.2 of \cite{Kl11}. Hence, we
omit the proof.

\textbf{Lemma 2.5}. \emph{Let the function }$r\left( k\right) $\emph{\ be
analytic for all }$k\in \mathbb{R}.$ \emph{Then the function }$\left\vert
r\left( k\right) \right\vert $\emph{\ can be uniquely determined for all }$%
k\in \mathbb{R}$ \emph{by the values of }$\left\vert r\left( k\right)
\right\vert $ \emph{\ for }$k\in \left( a,b\right) $\emph{.}

\textbf{Proof}. We have $\left\vert r\left( k\right) \right\vert
^{2}=r\left( k\right) \overline{r}\left( k\right) .$ Both $r\left( k\right) $
and $\overline{r}\left( k\right) $ are analytic functions of the real
variable $k$. $\square $

\textbf{Lemma 2.6}. \emph{Consider two finite sets of non-negative integers }%
$\left\{ p_{j_{1}}\right\} _{j_{1}=1}^{N_{1}}$\emph{\ and }$\left\{
q_{j_{2}}\right\} _{j_{2}=1}^{N_{2}}.$\emph{\ Also, consider two sets of
complex numbers }$\left\{ d_{j_{1}}\right\} _{j_{1}=1}^{N_{1}}\subset \left( 
\overline{\mathbb{C}}_{+}\diagdown \mathbb{R}\right) $\emph{\ and }$\left\{
s_{j_{2}}\right\} _{j_{2}=1}^{N_{2}}\subset \left( \overline{\mathbb{C}}%
_{+}\diagdown \mathbb{R}\right) .$\emph{\ Assume that there exist two sets
of complex numbers }$\left\{ \alpha _{j_{1}}\right\} _{j_{1}=1}^{N_{1}}$%
\emph{\ and }$\left\{ \beta _{j_{2}}\right\} _{j_{2}=1}^{N_{2}}$\emph{\ such
that}%
\begin{equation}
\dsum\limits_{j_{1}=1}^{N_{1}}\alpha _{j_{1}}t^{p_{j_{1}}}\exp \left( -i%
\overline{d}_{j_{1}}t\right) =\dsum\limits_{j_{1}=1}^{N_{1}}\beta
_{j_{2}}t^{q_{j_{2}}}\exp \left( -i\overline{s}_{j_{2}}t\right) ,\forall
t\geq 0.  \label{2.8}
\end{equation}%
\emph{Then }$N_{1}=N_{2}=N$\emph{\ and numbers involved in (\ref{2.8}) can
be re-numbered in such a way that }%
\begin{equation*}
\alpha _{j}=\beta _{j},p_{j}=q_{j},d_{j}=s_{j},\forall j=1,...,N.
\end{equation*}

\textbf{Proof}. Let $n\geq 0$ be an integer and let the number $a\in $ $%
\overline{\mathbb{C}}_{+}\diagdown \mathbb{R}.$ Consider the function $%
f_{n,a}\left( t\right) ,$%
\begin{equation}
f_{n,a}\left( t\right) =H\left( t\right) t^{n}\exp \left( -i\overline{a}%
t\right) .  \label{2.9}
\end{equation}%
Let $F_{n,a}\left( k\right) $ be the Fourier transform of $f_{n,a}\left(
t\right) ,$%
\begin{equation}
F_{n,a}\left( k\right) =\mathcal{F}\left( f_{n,a}\right) \left( k\right)
=\dint\limits_{-\infty }^{\infty }f_{n,a}\left( t\right) \exp \left(
ikt\right) dt.  \label{2.10}
\end{equation}%
Then the elementary calculation shows that%
\begin{equation}
F_{n,a}\left( k\right) =\frac{\left( -1\right) ^{n}n!}{i^{n+1}}\cdot \frac{1%
}{\left( k-\overline{a}\right) ^{n+1}}.  \label{2.11}
\end{equation}%
Hence, multiplying both sides of (\ref{2.8}) by the Heaviside function $%
H\left( t\right) $ and\emph{\ }applying then the operator $\mathcal{F}$ to
both sides of the resulting equality, we obtain for all $k\in \mathbb{R}$%
\begin{equation}
\dsum\limits_{j_{1}=1}^{N_{1}}\alpha _{j_{1}}\frac{\left( -1\right)
^{p_{j_{1}}}\left( p_{j_{1}}\right) !}{i^{p_{j_{1}}+1}}\cdot \frac{1}{\left(
k-\overline{d}_{j_{1}}\right) ^{p_{j_{1}}+1}}=\dsum\limits_{j_{2}=1}^{N_{2}}%
\beta _{j_{2}}\frac{\left( -1\right) ^{q_{j_{2}}}\left( q_{j_{2}}\right) !}{%
i^{q_{j_{2}}+1}}\cdot \frac{1}{\left( k-\overline{s}_{j_{2}}\right)
^{q_{j_{2}}+1}}.  \label{2.12}
\end{equation}%
Since (\ref{2.12}) is valid for all $k\in \mathbb{R}$ we can analytically
continue both sides of (\ref{2.12}) in $\mathbb{C}$ and obtain then
meromorphic functions in both sides of (\ref{2.12}). In other words, (\ref%
{2.12}) is valid for all $k\in \mathbb{C},$ except of poles. The rest of the
proof is obvious. $\square $

\section{Proof of Theorem 1}

\label{sec:3}

Consider two arbitrary points $y\in S,x\in B_{\omega }\left( y\right) ,x\neq
y$ mentioned in the formulation of this theorem. Denote $\psi \left(
k\right) =u\left( x,y,k\right) .$ Suppose that there exist two functions 
\begin{equation}
\psi _{1}\left( k\right) =u_{1}\left( x,y,k\right) ,\psi _{2}\left( k\right)
=u_{2}\left( x,y,k\right)  \label{3.0}
\end{equation}
such that $\left\vert \psi _{1}\left( k\right) \right\vert =\left\vert \psi
_{2}\left( k\right) \right\vert =f\left( x,y,k\right) ,k\in \left(
a,b\right) .$ Then by Lemma 2.5%
\begin{equation}
\left\vert \psi _{1}\left( k\right) \right\vert =\left\vert \psi _{2}\left(
k\right) \right\vert =f\left( x,y,k\right) ,\forall k\in \mathbb{R}.
\label{3.1}
\end{equation}%
By Lemma 2.1 functions $\psi _{1}\left( k\right) $ and $\psi _{2}\left(
k\right) $ can be analytically continued in the half plane $\mathbb{C}%
_{m}\subset \mathbb{C}.$ Also, recall that the upper half plane $\mathbb{C}%
_{+}\subset \mathbb{C}_{m}.$

Below we count each zero of any of two functions $\psi _{1}\left( k\right) $
and $\psi _{2}\left( k\right) $ as many times as its multiplicity is. First,
we show that real zeros of functions $\psi _{1}\left( k\right) $ and $\psi
_{2}\left( k\right) $ coincide. Let $\left\{ \alpha _{j}^{\left( 1\right)
}\right\} _{j=1}^{r_{1}}$ and $\left\{ \alpha _{s}^{\left( 2\right)
}\right\} _{s=1}^{r_{2}}$ be all real zeros of the function $\psi _{1}\left(
k\right) $ and $\psi _{2}\left( k\right) $ respectively. Let $\alpha
_{1}^{\left( 1\right) }$ be the real zero of $\psi _{1}\left( k\right) $ of
the multiplicity $n\geq 1.$ Also let $\alpha _{1}^{\left( 1\right) }$ be the
zero of the function $\psi _{2}\left( k\right) $ of the multiplicity $m\geq
0.$ Then 
\begin{equation}
\psi _{1}\left( k\right) =\left( k-\alpha _{1}^{\left( 1\right) }\right) ^{n}%
\widehat{\psi }_{1}\left( k\right) ,\psi _{2}\left( k\right) =\left(
k-\alpha _{1}^{\left( 1\right) }\right) ^{m}\widehat{\psi }_{2}\left(
k\right) ,  \label{3.2}
\end{equation}%
\begin{equation}
\widehat{\psi }_{1}\left( \alpha _{1}^{\left( 1\right) }\right) \neq 0,%
\widehat{\psi }_{2}\left( \alpha _{1}^{\left( 1\right) }\right) \neq 0.
\label{3.3}
\end{equation}%
Hence, by (\ref{3.1}) and (\ref{3.2}) 
\begin{equation}
\left\vert k-\alpha _{1}^{\left( 1\right) }\right\vert ^{n}\left\vert 
\widehat{\psi }_{1}\left( k\right) \right\vert =\left\vert k-\alpha
_{1}^{\left( 1\right) }\right\vert ^{m}\left\vert \widehat{\psi }_{2}\left(
k\right) \right\vert ,\forall k\in \mathbb{R}.  \label{3.4}
\end{equation}%
Let, for example $n>m.$ Dividing both sides of (\ref{3.4}) by $\left\vert
k-\alpha _{1}^{\left( 1\right) }\right\vert ^{m},$ we obtain%
\begin{equation*}
\left\vert \widehat{\psi }_{2}\left( k\right) \right\vert =\left\vert
k-\alpha _{1}^{\left( 1\right) }\right\vert ^{n-m}\left\vert \widehat{\psi }%
_{1}\left( k\right) \right\vert ,\forall k\in \mathbb{R}.
\end{equation*}%
Hence, $\widehat{\psi }_{2}\left( \alpha _{1}^{\left( 1\right) }\right) =0,$
which, however, contradicts to (\ref{3.3}). Hence, we have proven that real
zeros of both functions $\psi _{1}\left( k\right) $ and $\psi _{2}\left(
k\right) $ coincide. Let the set of real zeros of each of these functions be 
$\left\{ \alpha _{j}\right\} _{j=1}^{r}.$

Consider now complex zeros of functions $\psi _{1}\left( k\right) $ and $%
\psi _{2}\left( k\right) $ in the upper half plane $\mathbb{C}_{+}.$ By
Lemma 2.3 each of these two functions has at most a finite number of zeros
in $\mathbb{C}_{+}.$ Let $\left\{ a_{j}\right\} _{j=1}^{l_{1}}\subset 
\mathbb{C}_{+}$ and $\left\{ b_{s}\right\} _{s=1}^{l_{2}}\subset \mathbb{C}%
_{+}$ be those zeros of functions $\psi _{1}\left( k\right) $ and $\psi
_{2}\left( k\right) $ respectively. Consider functions $\mu _{1}\left(
k\right) $ and $\mu _{2}\left( k\right) $ defined as%
\begin{equation}
\mu _{1}\left( k\right) =\prod\limits_{j=1}^{r}\left( \frac{1}{k-\alpha _{j}}%
\right) \cdot \prod\limits_{j=1}^{l_{1}}\left( \frac{k-\overline{a}_{j}}{%
k-a_{j}}\right) \psi _{1}\left( k\right) ,k\in \mathbb{C}_{+},  \label{3.5}
\end{equation}%
\begin{equation}
\mu _{2}\left( k\right) =\prod\limits_{j=1}^{r}\left( \frac{1}{k-\alpha _{j}}%
\right) \cdot \prod\limits_{j=1}^{l_{2}}\left( \frac{k-\overline{b}_{j}}{%
k-b_{j}}\right) \psi _{2}\left( k\right) ,k\in \mathbb{C}_{+}.  \label{3.6}
\end{equation}%
Then 
\begin{equation}
\mu _{1}\left( k\right) \neq 0,\mu _{2}\left( k\right) \neq 0\text{ for }%
k\in \mathbb{C}_{+}\cup \mathbb{R}.  \label{3.7}
\end{equation}%
Furthermore, since 
\begin{equation*}
\left\vert \frac{k-\overline{a}}{k-a}\right\vert =1,\forall k\in \mathbb{R}%
,\forall a\in \mathbb{C},
\end{equation*}%
then (\ref{3.1}), (\ref{3.5}) and (\ref{3.6}) imply that%
\begin{equation}
\left\vert \mu _{1}\left( k\right) \right\vert =\left\vert \mu _{2}\left(
k\right) \right\vert ,\forall k\in \mathbb{R}.  \label{3.8}
\end{equation}

To apply Lemma 2.4, we now should show that the asymptotic behavior of
functions $\psi _{1}\left( k\right) $ and $\psi _{2}\left( k\right) $ for $%
\left\vert k\right\vert \rightarrow \infty ,k\in \mathbb{C}_{+}$ is the
same. It is natural to use formula (\ref{2.6}), which would be similar to 
\cite{KR1}. However, since functions $\psi _{1}\left( k\right) $ and $\psi
_{2}\left( k\right) $ supposedly correspond to two different coefficients $%
c_{1}\left( x\right) $ and $c_{2}\left( x\right) ,$ then they generate
different pairs of functions $\tau _{1}\left( x,y\right) ,A_{1}\left(
x,y\right) $ and $\tau _{2}\left( x,y\right) ,A_{2}\left( x,y\right) .$
This, in turn generates two different asymptotic behaviors in (\ref{2.6}).

Nevertheless, we still can use (\ref{2.6}). Indeed, since $y\in S,x\in
B_{\omega }\left( y\right) $ and also $B_{\omega }\left( y\right) \cap 
\overline{\Omega }=\varnothing ,$ then $c\left( x\right) =1$ in $B_{\omega
}\left( y\right) .$ Hence, $\tau \left( x,y\right) =\left\vert
x-y\right\vert $ for $x\in B_{\omega }\left( y\right) .$ Furthermore, by (%
\ref{1.91}) $J\left( x,y\right) =1$ for $x\in B_{\omega }\left( y\right) .$
Hence, by (\ref{2.30})%
\begin{equation*}
A\left( x,y\right) =\frac{1}{4\pi \left\vert x-y\right\vert },\text{ }x\in
B_{\omega }\left( y\right) .
\end{equation*}%
Hence, (\ref{2.6}) implies that%
\begin{equation*}
u(x,y,k)=\frac{\exp \left( ik\left\vert x-y\right\vert \right) }{4\pi
\left\vert x-y\right\vert }\left( 1+O\left( \frac{1}{k}\right) \right)
,\left\vert k\right\vert \rightarrow \infty ,k\in \mathbb{C}_{m},y\in S,x\in
B_{\omega }\left( y\right) .
\end{equation*}%
Hence, 
\begin{equation}
\psi _{j}\left( k\right) =\frac{\exp \left( ik\left\vert x-y\right\vert
\right) }{4\pi \left\vert x-y\right\vert }\left( 1+O\left( \frac{1}{k}%
\right) \right) ,\left\vert k\right\vert \rightarrow \infty ,k\in \mathbb{C}%
_{m},j=1,2.  \label{3.9}
\end{equation}%
Hence, using (\ref{3.5}), (\ref{3.6}) and (\ref{3.9}), we obtain 
\begin{equation}
\mu _{j}\left( k\right) =\frac{1}{k^{r}}\frac{\exp \left( ik\left\vert
x-y\right\vert \right) }{4\pi \left\vert x-y\right\vert }\left( 1+O\left( 
\frac{1}{k}\right) \right) ,\left\vert k\right\vert \rightarrow \infty ,k\in 
\mathbb{C}_{m},j=1,2.  \label{3.10}
\end{equation}%
Hence, (\ref{3.7}), (\ref{3.8}) and (\ref{3.10}) imply that we can apply
Lemma 2.4 to functions $\mu _{1}\left( k\right) $ and $\mu _{2}\left(
k\right) .$ We obtain then%
\begin{equation}
\mu _{1}\left( k\right) =\mu _{2}\left( k\right) ,k\in \mathbb{C}_{+}\cup 
\mathbb{R}.  \label{3.11}
\end{equation}

Using (\ref{3.5}), (\ref{3.6}) and (\ref{3.11}), we obtain%
\begin{equation*}
\prod\limits_{j=1}^{l_{1}}\left( \frac{k-\overline{a}_{j}}{k-a_{j}}\right)
\psi _{1}\left( k\right) =\prod\limits_{j=1}^{l_{2}}\left( \frac{k-\overline{%
b}_{j}}{k-b_{j}}\right) \psi _{2}\left( k\right) ,k\in \mathbb{R}.
\end{equation*}%
Or, equivalently,%
\begin{equation}
\prod\limits_{j=1}^{l_{2}}\left( \frac{k-b_{j}}{k-\overline{b}_{j}}\right)
\psi _{1}\left( k\right) =\prod\limits_{j=1}^{l_{1}}\left( \frac{k-a_{j}}{k-%
\overline{a}_{j}}\right) \psi _{2}\left( k\right) ,k\in \mathbb{R}.
\label{3.12}
\end{equation}

We now want to apply the operator $\mathcal{F}^{-1}$ of the inverse Fourier
transform (\ref{2.10}) to both sides of (\ref{3.12}). We rewrite (\ref{3.12}%
) as%
\begin{equation}
\psi _{1}\left( k\right) +\left( \prod\limits_{j=1}^{l_{2}}\frac{k-b_{j}}{k-%
\overline{b}_{j}}-1\right) \psi _{1}\left( k\right) =\psi _{2}\left(
k\right) +\left( \prod\limits_{j=1}^{l_{1}}\frac{k-a_{j}}{k-\overline{a}_{j}}%
-1\right) \psi _{2}\left( k\right) ,k\in \mathbb{R}.  \label{3.13}
\end{equation}%
Consider $\mathcal{F}^{-1}\left( \psi _{1}\right) $ and $\mathcal{F}%
^{-1}\left( \psi _{2}\right) .$ By (\ref{2.5}) and (\ref{2.50})%
\begin{equation}
\mathcal{F}^{-1}\left( \psi _{1}\right) =v_{1}\left( x,y,t\right) ,\mathcal{F%
}^{-1}\left( \psi _{2}\right) =v_{2}\left( x,y,t\right) ,  \label{3.14}
\end{equation}%
where $v_{1}\left( x,y,t\right) $ and $v_{2}\left( x,y,t\right) $ are
solutions of the problem (\ref{2.1}) and (\ref{2.2}) with two different
functions $c_{1}\left( x\right) $ and $c_{2}\left( x\right) $ which
correspond to functions $u_{1}\left( x,y,k\right) $ and $u_{2}\left(
x,y,k\right) $ respectively. Let $d=dist\left( y,\partial \Omega \right) .$
Then $d>\omega .$ On the other hand, since $x\in B_{\omega }\left( y\right)
, $ then $\left\vert x-y\right\vert <\omega <d.$ Let $t\in \left( 0,d\right)
.$ Since $c\left( x^{\prime }\right) =1,\forall x^{\prime }\in B_{\omega
}\left( y\right) ,$ then for $j=1,2$%
\begin{equation}
\partial _{t}^{2}v_{j}=\Delta v_{j}+\delta \left( x-y,t\right) ,x\in
B_{\omega }\left( y\right) ,t\in \left( 0,d\right) ,  \label{3.140}
\end{equation}%
\begin{equation}
v_{j}\mid _{t<0}\equiv 0.  \label{3.141}
\end{equation}%
Hence, it follows from the method of energy estimates that for $j=1,2$ 
\begin{equation}
v_{j}\left( x,y,t\right) =\frac{\delta \left( t-\left\vert x-y\right\vert
\right) }{4\pi \left\vert x-y\right\vert },x\in B_{\omega }\left( y\right)
,t\in \left( 0,d\right) .  \label{3.142}
\end{equation}%
Hence, using (\ref{3.14}) and (\ref{3.142}), we obtain 
\begin{equation}
\mathcal{F}^{-1}\left( \psi _{1}\right) =\mathcal{F}^{-1}\left( \psi
_{2}\right) =\frac{\delta \left( t-\left\vert x-y\right\vert \right) }{4\pi
\left\vert x-y\right\vert },t\in \left( 0,d\right) .  \label{3.15}
\end{equation}

Consider now functions $w_{1}\left( k\right) ,w_{2}\left( k\right) $ defined
as%
\begin{equation*}
w_{1}\left( k\right) =\prod\limits_{j=1}^{l_{1}}\frac{k-a_{j}}{k-\overline{a}%
_{j}}-1,
\end{equation*}%
\begin{equation*}
w_{2}\left( k\right) =\prod\limits_{j=1}^{l_{2}}\frac{k-b_{j}}{k-\overline{b}%
_{j}}-1.
\end{equation*}%
They can be rewritten as%
\begin{equation*}
w_{1}\left( k\right) =P_{1}\left( k\right) \prod\limits_{j=1}^{l_{1}}\frac{1%
}{k-\overline{a}_{j}},
\end{equation*}%
\begin{equation*}
w_{2}\left( k\right) =P_{2}\left( k\right) \prod\limits_{j=1}^{l_{2}}\frac{1%
}{k-\overline{b}_{j}},
\end{equation*}%
where $P_{1}\left( k\right) $ is the polynomial of the degree less than $%
l_{1}$ and $P_{2}\left( k\right) $ is the polynomial of the degree less than 
$l_{2}.$ Using the partial fraction expansion, we obtain%
\begin{equation*}
w_{1}\left( k\right) =\dsum\limits_{j_{1}=1}^{l_{1}^{\prime }}\frac{Y_{j_{1}}%
}{\left( k-\overline{a}_{j_{1}}\right) ^{p_{j_{1}}}},
\end{equation*}%
\begin{equation*}
w_{2}\left( k\right) =\dsum\limits_{j_{2}=1}^{l_{2}^{\prime }}\frac{Z_{j_{2}}%
}{\left( k-\overline{b}_{j_{2}}\right) ^{q_{j_{1}}}},
\end{equation*}%
where $l_{1}^{\prime }\leq l_{1},l_{2}^{\prime }\leq l_{2}$, the sets $%
\left\{ \overline{a}_{j_{1}}\right\} _{j_{1}=1}^{l_{1}^{\prime }}\subseteq
\left\{ \overline{a}_{j}\right\} _{j=1}^{l_{1}},\left\{ \overline{b}%
_{j_{2}}\right\} _{j_{2}=1}^{l_{2}^{\prime }}\subseteq \left\{ \overline{b}%
_{j}\right\} _{j=1}^{l_{2}}$ and $Y_{j_{1}}$, $Z_{j_{2}}$ are some complex
numbers. \ 

We now apply the operator $\mathcal{F}^{-1}$ to functions $w_{1}\left(
k\right) ,w_{2}\left( k\right) .$ By (\ref{2.9})-(\ref{2.11})%
\begin{equation*}
\mathcal{F}^{-1}\left( \frac{Y_{j_{1}}}{\left( k-\overline{a}_{j_{1}}\right)
^{p_{j_{1}}}}\right) =H\left( t\right) \frac{\left( -1\right)
^{p_{j_{1}}-1}i^{p_{j_{1}}}Y_{j_{1}}}{\left( p_{j_{1}}-1\right) !}%
t^{p_{j_{1}}-1}\exp \left( -i\overline{a}_{j_{1}}t\right) .
\end{equation*}%
Hence,%
\begin{equation*}
\mathcal{F}^{-1}\left( w_{1}\right) =Q_{1}\left( t\right) ,\mathcal{F}%
^{-1}\left( w_{2}\right) =Q_{2}\left( t\right) ,
\end{equation*}%
\begin{equation}
Q_{1}\left( t\right) =H\left( t\right) \dsum\limits_{j_{1}=1}^{l_{1}^{\prime
}}\frac{\left( -1\right) ^{p_{j_{1}}-1}i^{p_{j_{1}}}Y_{j_{1}}}{\left(
p_{j_{1}}-1\right) !}t^{p_{j_{1}}-1}\exp \left( -i\overline{a}%
_{j_{1}}t\right) ,  \label{3.17}
\end{equation}%
\begin{equation}
Q_{2}\left( t\right) =H\left( t\right) \dsum\limits_{j_{2}=1}^{l_{2}^{\prime
}}\frac{\left( -1\right) ^{q_{j_{2}}-1}i^{q_{j_{2}}}Z_{j_{2}}}{\left(
q_{j_{2}}-1\right) !}t^{p_{j_{1}}-1}\exp \left( -i\overline{b}%
_{j_{2}}t\right) .  \label{3.18}
\end{equation}%
We are ready now to apply the operator $\mathcal{F}^{-1}$ to both parts of (%
\ref{3.13}). Using the convolution theorem for the Fourier transform, (\ref%
{3.14})-(\ref{3.141}), (\ref{3.17}) and (\ref{3.18}), we obtain%
\begin{equation}
v_{1}\left( x,y,t\right) +\dint\limits_{0}^{t}v_{1}\left( x,y,t-\tau \right)
Q_{1}\left( \tau \right) d\tau =v_{2}\left( x,y,t\right)
+\dint\limits_{0}^{t}v_{2}\left( x,y,t-\tau \right) Q_{2}\left( \tau \right)
d\tau .  \label{3.19}
\end{equation}%
Let $t\in \left( 0,d\right) .$ Then, using (\ref{3.142}), we obtain that (%
\ref{3.19}) becomes%
\begin{equation*}
\dint\limits_{0}^{t}\delta \left( t-\tau -\left\vert x-y\right\vert \right)
Q_{1}\left( \tau \right) d\tau =\dint\limits_{0}^{t}\delta \left( t-\tau
-\left\vert x-y\right\vert \right) Q_{2}\left( \tau \right) d\tau ,\forall
t\in \left( 0,d\right) .
\end{equation*}%
This is equivalent with%
\begin{equation}
Q_{1}\left( t-\left\vert x-y\right\vert \right) =Q_{2}\left( t-\left\vert
x-y\right\vert \right) ,\forall t\in \left( 0,d\right) .  \label{3.20}
\end{equation}%
By (\ref{3.17}) and (\ref{3.18}) both functions $Q_{1}\left( t\right)
,Q_{2}\left( t\right) $ are analytic functions of the real variable $t>0.$
Hence, since $d>\left\vert x-y\right\vert ,$ then (\ref{3.20}) is valid for
all $t>\left\vert x-y\right\vert .$ Hence, denoting $\widetilde{t}%
=t-\left\vert x-y\right\vert ,$ we obtain 
\begin{equation}
Q_{1}\left( \widetilde{t}\right) =Q_{2}\left( \widetilde{t}\right) ,\forall 
\widetilde{t}>0.  \label{3.21}
\end{equation}%
Finally applying Lemma 2.6 to (\ref{3.21}), we obtain that $l_{1}=l_{2}:=l$
and sets $\left\{ a_{1},a_{2},...,a_{l}\right\} =\left\{
b_{1},b_{2},...,b_{l}\right\} .$ Thus, by (\ref{3.12}) $\psi _{1}\left(
k\right) =\psi _{2}\left( k\right) $ for $k\in \mathbb{R}.$ This and (\ref%
{3.0}) lead to 
\begin{equation}
u_{1}\left( x,y,k\right) =u_{2}\left( x,y,k\right) ,k\in \mathbb{R}.
\label{3.22}
\end{equation}%
So, (\ref{3.22}) finalizes the proof of the first part of Theorem 1.

We now prove that the coefficient $c\left( x\right) $ is determined
uniquely. The equality (\ref{3.22}) is valid for that fixed pair $y\in
S,x\in B_{\omega }\left( y\right) .$ Since $x$ is an arbitrary point of $%
B_{\omega }\left( y\right) ,$ then (\ref{3.22}) holds for all $x\in
B_{\omega }\left( y\right) .$ We now return to the original equation (\ref%
{1.4}). We have proven that for each source position $y\in S$ the function $%
u\left( x,y,k\right) $ is determined uniquely for points $x\in B_{\omega
}\left( y\right) $ and for all $k$ $\in \mathbb{R}.$ Fix a point $y\in S.$
Since by (\ref{1.1}) and (\ref{1.3}) $c\left( x\right) =1$ outside of the
domain $\Omega $ and since $B_{\omega }\left( y\right) \cap \overline{\Omega 
}=\varnothing ,$ then the well known theorem about the uniqueness of the
continuation of the solution of an arbitrary elliptic equation of the second
order (see, e.g. Chapter 4 of \cite{LRS}) implies that the function $u\left(
x,y,k\right) $ is determined uniquely for all $x\notin \overline{\Omega }.$
Next, since $y\in S$ is an arbitrary point, then the function $u\left(
x,y,k\right) $ is uniquely determined for all $y\in S$ and for all $x\notin 
\overline{\Omega }.$ By (\ref{2.5}) and (\ref{2.50}) this means, in turn
that the function $v\left( x,y,t\right) $ is uniquely determined for all $%
y\in S,x\notin \overline{\Omega },t>0.$ This and (\ref{2.3}) imply that the
following function $s\left( x,y\right) $ is known%
\begin{equation}
s\left( x,y\right) =\tau \left( x,y\right) ,\forall x,y\in S.  \label{3.23}
\end{equation}%
Recall that the function $\tau \left( x,y\right) $ satisfies the eikonal
equation (\ref{1.9}) with condition (\ref{1.900}). The problem of the
determination of the function $c\left( x\right) $ from the function $\tau
\left( x,y\right) $ known for all $x,y\in S$ is called Inverse Kinematic
Problem \cite{LRS,R1,R2}. In the 3-D case uniqueness of this problem was
proven in Theorem 3.4 of Chapter 3 of \cite{R2}. Therefore the uniqueness of
the determination of the unknown coefficient $c\left( x\right) $ is
established. $\square $

\end{document}